\documentclass[aps,prb,fleqn,12pt,showpacs,superscriptaddress]{revtex4}
\usepackage{epsfig,color}
\usepackage{graphicx}
\usepackage{pdfpages}
\usepackage{braket}
\usepackage{amssymb,amsmath}
\usepackage{amsfonts} 
\usepackage{hyperref} 

\begin{document}
\title{Intra- and inter-orbital correlated electron spin dynamics in $\rm Sr_2 Ir O_4$: spin-wave gap and spin-orbit exciton}
\author{Shubhajyoti Mohapatra}
\affiliation{Department of Physics, Indian Institute of Technology, Kanpur - 208016, India}
\author{Avinash Singh}
\affiliation{Department of Physics, Indian Institute of Technology, Kanpur - 208016, India}
\email{shubhajm@iitk.ac.in}
\date{\today} 
\begin{abstract}
Transformation of Coulomb interaction terms to the pseudo-orbital basis constituted by $J=1/2$ and $3/2$ states arising from spin-orbit coupling provides a versatile tool. This formalism is applied to investigate magnetic anisotropy effects on low-energy spin-wave excitations as well as high-energy spin-orbit exciton modes in $\rm Sr_2 Ir O_4$. The Hund's coupling term explictly yields easy-plane anisotropy, resulting in gapless (in-plane) and gapped (out-of-plane) modes, in agreement with recent resonant inelastic x-ray scattering (RIXS) measurements. The collective mode of inter-orbital, spin-flip, particle-hole excitations with appropriate interaction strengths and renormalized spin-orbit gap yields two well-defined propagating spin-orbit exciton modes, with energy scale and dispersion in good agreement with RIXS studies. 
\end{abstract}
\maketitle

\newpage
\section{Introduction}
The iridium based transition-metal oxides exhibiting novel $J$=1/2 Mott insulating states have attracted considerable interest in recent years in view of their potential for hosting collective quantum states such as quantum spin liquids, topological orders, and high-temperature superconductors.\cite{krempa_AR_2014} The effective $J$=1/2 antiferromagnetic (AFM) insulating state in iridates arises from a novel interplay between crystal field, spin-orbit coupling and intermediate Coulomb correlations. Exploration of the emerging quantum states in the iridate compounds therefore involves investigation of the correlated spin-orbital entangled electronic states and related magnetic properties. 

Among the iridium compounds, the quasi-two-dimensional (2D) square-lattice perovskite-structured iridate $\rm Sr_2IrO_4$ is of special interest as the first spin-orbit Mott insulator to be identified and because of its structural and physical similarity with $\rm La_2CuO_4$.\cite{rau_AR_2016,bertinshaw_AR_2018} It exhibits canted AFM ordering of the pseudospins below N\'{e}el temperature $T_{\rm N} \approx 240$ K. The canting of the in-plane magnetic moments tracks the staggered $\rm IrO_6$ octahedral rotations about the $c$ axis. The effectively single (pseudo) orbital ($J$=1/2) nature of this Mott insulator has motivated intensive finite doping studies aimed at inducing the superconducting state as in the cuprates.\cite{senthil_PRL_2011,kim3_SC_2014,torre_PRL_2015,kim4_NAT_2016,gretarsson_PRL_2016,chen_NATCOM_2018,
bhowal_JPCM_2018}

Recent technological advancements in resonant inelastic X-ray scattering (RIXS) have been instrumental in the elucidation of the pseudospin dynamics in $\rm Sr_2IrO_4$. In the first published data,\cite{kim1_PRL_2012} spectra along high-symmetry directions in the Brillouin zone reveal a single gapless spin-wave mode with a dispersion of $\sim$200 meV, indicating isotropic nature of pseudo-spin interactions. In subsequent investigations of both parent and electron-doped compounds, the limited energy resolution of RIXS could not also resolve any spin-wave gap.\cite{liu_PRB_2016,gretarsson_PRL_2016} However, recent measurements conducted with improved energy resolution point to a partially resolved $\sim$30 meV spin-wave gap at the $\Gamma$ point,\cite{pincini_PRB_2017} which has been further resolved via high-resolution RIXS and inelastic neutron scattering (INS), both of which indicate another spin-wave gap between 2 to 3 meV at $(\pi, \pi)$.\cite{porras_arxiv_2018} These low-energy features correspond to different spin-wave modes associated with basal-plane and out-of-plane fluctuations, indicating the presence of anisotropic spin interactions. 

In addition to spin-wave modes, RIXS experiments have also revealed a high-energy dispersive feature in the energy range 0.4-0.8 eV. Attributed to  electron-hole pair excitations across the spin-orbit gap between the $J$=1/2 and 3/2 bands, this distinctive mode is referred to as the spin-orbit exciton.\cite{kim1_PRL_2012,kim_NATCOMM_2014,lu_PRB_2018,kim2_PRL_2012,igarashi_PRB_2014} Unusual magnetism has been predicted in recent theoretical investigations for $5d^4$ and $5d^5$ systems arising from the condensation of spin-orbit excitons.\cite{khaliullin_PRL_2013,sato_PRB_2015,valenti_PRL_2017,kaushal_arxiv_2019} 

The anisotropic magnetic interactions such as the pseudo-dipolar (PD) and Dzyaloshinskii-Moriya (DM) terms within the effective $J$=1/2 spin model for $\rm Sr_2IrO_4$ account for the canted AF state, but do not yield true magnetic anisotropy and spin-wave gap as the relevant spin-dependent hopping term can be gauged away.\cite{jackeli_PRL_2009,senthil_PRL_2011,iridate1_PRB_2017} The easy basal-plane anisotropy has been proposed to arise from the Hund's coupling term when virtual excitations to $J$=3/2 states are included.\cite{jackeli_PRL_2009,igarashi_PRB_2013,perkins_PRB_2014,vale_PRB_2015} Recently, the pseudo-spin-lattice coupling has been proposed to account for the structural orthorhombicity, the easy-axis anisotropy within the basal plane, and alignment of moments along the crystallographic direction.\cite{liu_arxiv_2018} However, electron itineracy, finite mixing between $J$=1/2 and 3/2 sectors, and weak correlation effects play key roles in explaining the magnetic properties of $\rm Sr_2IrO_4$, and thus put limitations on these phenomenological spin models. 

In terms of multi-orbital itinerant-electron approaches, collective magnetic excitations were studied within the Hartree-Fock (HF) and the random phase approximation (RPA).\cite{igarashi_JPSJ_2014,iridate1_PRB_2017} In ref. [29], the spin-wave  mode was shown to be split into two branches - one gapless and the other gapped - and the spin-wave gap was explained in terms of the anisotropic exchange coupling attributed to the interplay between Hund's coupling and spin-orbit coupling. However, the crucial role of the finite magnetic moment in the nominally filled $J$=3/2 sector on the expression of the magnetic anisotropy was not studied. In ref. [24], the focus was on understanding the strong zone-boundary spin-wave dispersion, which was demonstrated as arising from finite-$U$ and finite-SOC effects.

Appreciable mixing between $J$=1/2 and 3/2 sectors, especially near the Fermi energy, has been shown in investigations of the pseudo-orbital-resolved electronic bands using the density functional theory (DFT) approach\cite{martins_PRL_2011,arita_PRL_2012,zhang_PRL_2013} and realistic three-orbital models.\cite{watanabe_PRL_2010,carter_PRB_2013,iridate1_PRB_2017,zhou_PRX_2017} Significant deviation from ideal fillings ($n_{1/2}$=1 and $n_{3/2}$=4) for the two sectors in $\rm Sr_2IrO_4$,\cite{martins_PRL_2011} and small magnetic moment in the $J$=3/2 sector have also been reported,\cite{carter_PRB_2013,iridate1_PRB_2017} implying break-down of the one-band ($J$=1/2) picture in real systems. Within a minimal extension of this picture which can provide a unified description of the observed high-energy features as discussed above, investigation of the coupling and excitations between the $J$=1/2 and $J$=3/2 sectors is therefore of particular interest.


In this paper, we therefore plan to investigate intra- and inter-orbital correlated-electron spin dynamics in $\rm Sr_2IrO_4$. Detailed comparison with RIXS data can provide experimental evidence of the several distinctive features associated with the rich interplay of spin-orbit coupling, Coulomb interaction, and realistic multi-orbital electronic band structure. These key features include: (i) dispersion of spin-wave and spin-orbit exciton modes, (ii) finite-$U$ and finite-SOC effects, (iii) mixing between $J$=1/2 and 3/2 sectors, (iv) Hund's-coupling-induced spin-rotation-symmetry-breaking and spin-wave gap, (v) correlation-induced spin-orbit gap renormalization, (vi) coupling between collective and single-particle excitations, and (vii) coupling between magnetic moments in the $J$=1/2 and 3/2 sectors. The structure of the paper is as follows. 

After a brief description of the three-orbital model and the pseudo-spin-orbital basis in Sec. II, the transformation of various Coulomb interaction terms from the original three-orbital basis to the pseudo-orbital basis is presented in Sec. III, explicitly showing  easy $x$-$y$ plane anisotropy resulting from the Hund's coupling term. Representation of the AFM state in the pseudo-orbital basis is discussed in Sec. IV. Formulation of the spin-wave propagator and the calculated dispersion showing the spin-wave gap are presented in Secs. V and VI. The spin-orbit gap renormalization due to the relative energy shift between the $J$=1/2 and 3/2 states arising from the density interaction terms is discussed in Sec. VII. The spin-orbit exciton as a resonant state formed by the correlated propagation of the inter-orbital, spin-flip, particle-hole excitation across the renormalized spin-orbit gap is investigated in Sec. VIII. Finally, conclusions are presented in Sec. IX.

\section{Three-orbital model: pseudo-orbital basis}
Due to large crystal-field splitting ($\sim$3 eV) in the $\rm Ir O_6$ octahedra, the low-energy physics in the $d^5$ iridates is effectively described by projecting out the empty ${\rm e_g}$ levels which are well above the $t_{\rm 2g}$ levels. Spin-orbit coupling (SOC) further splits the t$_{2g}$ states into (upper) $J$=1/2 doublet and (lower) $J$=3/2 quartet with an energy gap of $3\lambda/2$. Four of the five electrons fill the $J$=3/2 states, leaving one electron for the $J$=1/2 sector, rendering it magnetically active in the ground state. 

\begin{figure}
\vspace*{0mm}
\hspace*{0mm}
\psfig{figure=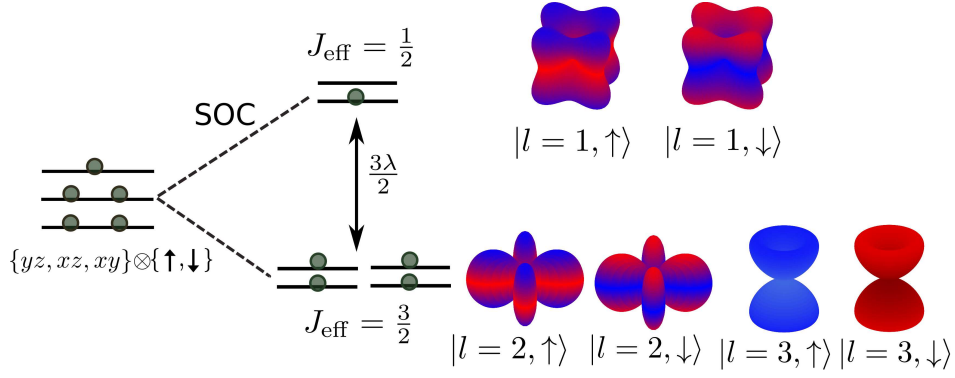,angle=0,width=125mm}
\vspace{-0mm}
\caption{The `pseudo-spin-orbital' energy level scheme for the three Kramers pairs along with their orbital shapes. The colors represent the weights of real spin $\uparrow$ (red) and $\downarrow$ (blue) in each pair.} 
\label{schematic}
\end{figure}

Corresponding to the three Kramers pairs above, we introduce three {\em pseudo orbitals} ($l=1,2,3$) with {\em pseudo spins} ($\tau= \uparrow,\downarrow$) each. The $|J,m_j \rangle$ and the corresponding $|l,\tau \rangle$ states have the form:
\begin{eqnarray}
\ket{l=1, \tau= \sigma} &=& \Ket{\frac{1}{2},\pm\frac{1}{2}} = \left [\Ket{yz,\bar{\sigma}} \pm i \Ket{xz,\bar{\sigma}} \pm \Ket{xy,\sigma}\right ] / \sqrt{3} \nonumber \\
\ket{l=2, \tau= \sigma} &=& \Ket{\frac{3}{2},\pm\frac{1}{2}} = \left [\Ket{yz,\bar{\sigma}} \pm i \Ket{xz,\bar{\sigma}} \mp 2 \Ket{xy,\sigma} \right ] / \sqrt{6} \nonumber \\
\ket{l=3, \tau= \bar{\sigma}} &=& \Ket{\frac{3}{2},\pm\frac{3}{2}} = \left [\Ket{yz,\sigma} \pm i \Ket{xz,\sigma}\right ] / \sqrt{2} 
\label{jmbasis}
\end{eqnarray}
where $\Ket{yz,\sigma}$, $\Ket{xz,\sigma}$, $\Ket{xy,\sigma}$ are the t$_{2g}$ states and the signs $\pm$ correspond to spins $\sigma = \uparrow/\downarrow$. The coherent superposition of different-symmetry $t_{\rm 2g}$ orbitals, with opposite spin polarization between $xz$/$yz$ and $xy$ levels implies spin-orbital entanglement, and also imparts unique extended 3D shape to the pseudo-orbitals $l=1, 2, 3$, as shown in Fig \ref{schematic}. 

Inverting the above transformation, we obtain the representation of the three-orbital basis states in terms of the pseudo-orbital basis states, given below in terms of the corresponding creation operators:

\begin{equation}
\begin{pmatrix}
a_{yz \sigma}^\dagger \\
  a_{xz \sigma}^\dagger \\
  a_{xy \overline{\sigma}}^\dagger
 \end{pmatrix}
 =
 \begin{pmatrix}
  \frac{1}{\sqrt{3}} & \frac{1}{\sqrt{6}} & \frac{1}{\sqrt{2}} \\
  \frac{i\sigma}{\sqrt{3}} & \frac{i\sigma}{\sqrt{6}} & \frac{-i\sigma}{\sqrt{2}} \\

 \frac{-\sigma}{\sqrt{3}} & \frac{\sqrt{2}\sigma}{\sqrt{3}} & 0
 \end{pmatrix}
 \begin{pmatrix}
  a_{1 \tau}^\dagger \\

  a_{2 \tau}^\dagger \\

  a_{3 \tau}^\dagger
\end{pmatrix}
\end{equation}
where, $\sigma = \uparrow/\downarrow$ and $\tau = \overline{\sigma}$.

Now, we consider the free part of the three-orbital model Hamiltonian including the SOC and band terms represented in the basis $(yz\sigma,xz\sigma,xy\bar{\sigma})$: 
\begin{equation}
\mathcal{H}_{\rm SO} + \mathcal{H}_{\rm band} = \sum_{{\bf k} \mu \sigma} \psi_{{\bf k} \mu \sigma}^{\dagger} \begin{pmatrix}
{\cal E}_{\bf k} ^{yz} & i \sigma\frac{\lambda}{2} + {\cal E}_{\bf k} ^{yz|xz} & -\sigma\frac{\lambda}{2} \\
- i \sigma\frac{\lambda}{2} + {\cal E}_{\bf k} ^{yz|xz} & {\cal E}_{\bf k} ^{xz} & i\frac{\lambda}{2} \\
-\sigma\frac{\lambda}{2} & - i\frac{\lambda}{2} & {\cal E}_{\bf k} ^{xy} \\
\end{pmatrix} \psi_{{\bf k} \mu \sigma} 
\label{three_orb_matrix}
\end{equation}  
where $\psi_{{\bf k} \mu \sigma}^{\dagger} = \left( a_{{\bf k} yz \sigma}^\dagger \; a_{{\bf k} xz \sigma}^\dagger \; a_{{\bf k} xy \bar \sigma}^\dagger\right)$, ${\cal E}_{\bf k} ^\mu$ are the band energies for the three orbitals $\mu$, and $\lambda$ is the SOC constant. The orbital mixing hopping term ${\cal E}^{yz|xz}_{\bf k}$ arises from the staggered $\rm IrO_6$ octahedral rotations in $\rm Sr_2IrO_4$.

Applying the transformation given in Eq. (2), the above Hamiltonian is transformed to the pseudo-orbital basis $|1,\tau = \uparrow,\downarrow\rangle $, $|2,\tau = \uparrow,\downarrow\rangle$, and $|3,\tau =\uparrow,\downarrow\rangle$. The orbital mixing hopping term ${\cal E}^{yz|xz}_{\bf k}$ leads to pseudo-spin-dependent terms in this basis, which breaks spin-rotation symmetry. However, these spin-dependent terms can be gauged away by a spin- and site-dependent unitary transformation,\cite{iridate1_PRB_2017} leaving the spin-independent form: $\mathcal{H}_{\rm SO} + \mathcal{H}_{\rm band} = \sum_{{\bf k} l m} {\cal E}_{\bf k} ^{lm} \psi_{{\bf k} l}^{\dagger} {1\!\!1} \psi_{{\bf k} m}$, which is invariant under the SU(2) transformation $\psi_{{\bf k}m} \rightarrow \psi_{{\bf k}m} ' = [U]\psi_{{\bf k}m}$ in pseudo-spin space.

In the above discussion, the two magnetic sublattices corresponding to the staggered magnetic order have not been included for compactness. The band term ${\cal E}_{\bf k}$ includes nearest-neighbor (NN) and next-nearest-neighbor (NNN) hopping terms etc., which therefore connect different or same magnetic sublattice(s), as will be discussed in Sec. IV. 

\section{Coulomb interaction terms in pseudo-orbital basis}

We consider the on-site Coulomb interaction terms:
\begin{eqnarray}
\mathcal{H}_{\rm int} &=& U\sum_{i,\mu}{n_{i\mu\uparrow}n_{i\mu\downarrow}} + U^\prime \sum_{i,\mu < \nu,\sigma} {n_{i\mu\sigma}n_{i\nu\overline{\sigma}}} + (U^\prime - J_{\mathrm H}) \sum_{i,\mu < \nu,\sigma}{n_{i\mu\sigma}n_{i\nu\sigma}} \nonumber\\ 
&&+ 
J_{\mathrm H} \sum_{i,\mu \ne \nu} ({a_{i \mu \uparrow}^{\dagger}a_{i \nu\downarrow}^{\dagger}a_{i \mu \downarrow} a_{i \nu \uparrow}} + {a_{i \mu \uparrow}^{\dagger} a_{i \mu\downarrow}^{\dagger}a_{i \nu \downarrow} a_{i \nu \uparrow}})
\end{eqnarray} 
in the three-orbital basis ($\mu,\nu = yz, xz, xy$), including the intra-orbital $(U)$ and inter-orbital $(U')$ density interaction terms, the Hund's coupling term $(J_{\rm H})$, and the pair hopping term $(J_{\rm H})$. Here $a_{i\mu\sigma}^{\dagger}$ and $a_{i\mu \sigma}$ are the creation and annihilation operators for site $i$, orbital $\mu$, spin $\sigma=\uparrow ,\downarrow$, and the density operator $n_{i \mu \sigma} = a_{i \mu \sigma}^\dagger a_{i\mu \sigma}$. 

Using the transformation from the three-orbital basis to the pseudo-orbital basis $(m,m'=1,2,3)$ described earlier, and keeping density as well as spin-flip interaction terms which are relevant for the present study, we obtain (for site $i$):
\begin{eqnarray}
\mathcal{H}_{\rm int} (i) &=& 
\frac{1}{2} \sum_{m, m^\prime, \tau, \tau^\prime} 
\mathcal{U}_{m m^\prime}^{\tau \tau^\prime} n_{m \tau} n_{m^\prime \tau^\prime} 
+ \left( \frac{U - U ^\prime}{3} \right) \sum_{\tau} {a_{1 \tau}^{\dagger}a_{2 \overline{\tau}}^{\dagger}a_{1 \overline{\tau}} a_{2 \tau}} \nonumber \\
&&+ 
\left( \frac{U - 2J_{\rm H} - U ^\prime}{6} \right) \sum_{\tau} \left( a_{2 \tau}^{\dagger}a_{3 \overline{\tau}}^{\dagger}a_{2 \overline{\tau}} a_{3 \tau} + 2 a_{3 \tau}^{\dagger}a_{1 \overline{\tau}}^{\dagger}a_{3 \overline{\tau}} a_{1 \tau}  \right)
\end{eqnarray}
where the transformed interaction matrices $\mathcal{U}_{m m^\prime}^{\tau \tau^\prime}$ in the new basis have the following form:
\begin{eqnarray}
\mathcal{U}_{m m^\prime}^{\tau \tau} &=& \left( 
\begin{array}{cccc}
0  & U^\prime & U^\prime - \frac{2}{3} J_{\rm H} \\
U^\prime & 0 & U^\prime-\frac{1}{3} J_{\rm H} \\
U^\prime-\frac{2}{3} J_{\rm H} & U^\prime-\frac{1}{3} J_{\rm H} & 0
\end{array}
\right), \nonumber \\
\mathcal{U}_{m m^\prime}^{\tau \overline{\tau}} &=& \left( 
\begin{array}{cccc}
\frac{1}{3} (U + 2 U^\prime)  & \frac{1}{3} (U + 2 U^\prime - 3 J_{\rm H}) & \frac{1}{3} (U + 2 U^\prime - J_{\rm H}) \\ 
\frac{1}{3} (U + 2 U^\prime - 3 J_{\rm H}) & \frac{1}{2} (U + U^\prime) & \frac{1}{6} (U + 5 U^\prime - 4 J_{\rm H}) \\
\frac{1}{3} (U + 2 U^\prime - J_{\rm H}) & \frac{1}{6} (U + 5 U^\prime - 4 J_{\rm H}) & \frac{1}{2} (U +  U^\prime) 
\end{array} 
\right)
\label{cal_u_eqns}
\end{eqnarray}
for pseudo-spins $\tau^\prime = \tau$ and $\tau^\prime = \overline{\tau}$, where $\tau = \uparrow, \downarrow$. Similar transformation to the $J$ basis has been discussed recently, focussing only on the density interaction terms.\cite{martins_JPCM_2017}

Using the spherical symmetry condition ($U^\prime$=$U$-$2J_{\mathrm H}$), the transformed interaction Hamiltonian can be written in terms of the local density and spin operators as:
\begin{eqnarray}
{\mathcal H}_{\rm int}(i) &=& \left( U - \frac{4}{3} J_{\rm H} \right) n_{1 \uparrow} n_{1 \downarrow} + \left( U - J_{\rm H} \right) \left[ n_{2 \uparrow}  n_{2 \downarrow} +  n_{3 \uparrow} n_{3 \downarrow} \right] \nonumber \\
&-& \frac{4}{3} J_{\rm H} {\bf S}_1 . {\bf S}_2 + 2 J_{\rm H} \left [ \mathcal{S}_{1}^z \mathcal{S}_{2}^z - \mathcal{S}_{1}^z \mathcal{S}_{3}^z \right] \nonumber \\
&+& \left( U-\frac{13}{6} J_{\rm H} \right) \left[ n_{1} n_{2} + n_{1} n_{3} \right] + 
\left( U-\frac{7}{3} J_{\rm H} \right) n_{2} n_{3} 
\label{h_int}
\end{eqnarray}
where the spin operator ${\bf S}_m = \psi_m ^\dagger \frac{\mbox{\boldmath $\tau$}}{2} \psi_m$ and the density operator $n_m = \psi_m ^\dagger {\bf 1} \psi_m = n_{m \uparrow} + n_{m \downarrow}$ in terms of the local pseudo-spin-orbital field operator $\psi_m ^\dagger = (a_{m\uparrow} ^\dagger \; a_{m \downarrow} ^\dagger )$. 



The Hubbard-like interaction terms ${\cal U}_m n_{m\uparrow} n_{m\downarrow} \sim -{\cal U}_m {\bf S}_m . {\bf S}_m$ are  invariant under pseudo-spin rotation, as is the Hund's-coupling-like term ${\bf S}_1.{\bf S}_2$. Furthermore, under the corresponding SU(2) transformation $\psi_m \rightarrow \psi_m ' = [U]\psi_m$, the total density terms $n_m$ are invariant. Therefore, the only interaction terms which break spin rotation symmetry and are thus responsible for magneto-crystalline anisotropy in $\rm Sr_2 Ir O_4$ are the $S_1 ^z S_2 ^z$ and $S_1 ^z S_3 ^z$ terms. As discussed earlier, the magnetically active sector is the nominally half-filled $m=1$ pseudo-orbital. Magnetic moments in the nominally doubly occupied $m=2,3$ orbitals are very small. As $S_2 ^z$ and $S_3 ^z$ are proportional to $S_1 ^z$ within a classical spin picture, the magnetic anisotropy terms can be written as $D (S_1 ^z)^2$, corresponding to an effective single-ion anisotropy. As shown below, we will find that $D>0$, indicating easy $x$-$y$ plane anisotropy.

\section{AF state: staggered field term}
We consider the ($\pi, \pi$) ordered AF state on the square lattice, focussing on the staggered field terms within the pseudo-orbital basis arising from the Hartree-Fock (HF) approximation of the various interaction terms in Eq. (\ref{h_int}). The charge terms corresponding to density condensates in this approximation will be discussed in Sec. VII. For general ordering direction with components {\boldmath $\Delta_l$}= $(\Delta_l ^x,\Delta_l ^y,\Delta_l ^z)$, the staggered field term for sector $l$ is given by:  
\begin{equation}
\mathcal{H}_{\rm sf} (l) = 
\sum_{{\bf k} s}  \psi_{{\bf k}ls}^{\dagger} 
\begin{pmatrix} -s \makebox{\boldmath $\tau . \Delta_l$}
\end{pmatrix} \psi_{{\bf k}ls} 
= \sum_{{\bf k} s} -s \psi_{{\bf k}ls}^{\dagger} 
\begin{pmatrix} \Delta_l ^z & \Delta_l ^x - i \Delta_l ^y \\
\Delta_l ^x + i \Delta_l ^y & -\Delta_l ^z  \\
\end{pmatrix} \psi_{{\bf k}ls}
\label{gen_ord_dirn} 
\end{equation}  
where $\psi_{{\bf k}ls} ^\dagger = (a_{{\bf k}ls\uparrow} ^\dagger \; \; a_{{\bf k}ls\downarrow} ^\dagger)$, $s=\pm 1$ for the two sublattices A/B, and the staggered field components $\Delta_{l=1,2,3} ^{x,y,z}$ are self-consistently determined from:
\begin{eqnarray}
2 \Delta^z _{1} &=& {\mathcal U}_1 m_1 ^z + 
\frac{2J_{\rm H}}{3} m_2 ^z + J_{\rm H} (m_3 ^z - m_2 ^z) \nonumber \\
2 \Delta^{x,y} _{1} &=& {\mathcal U}_1 m_1 ^{x,y} + \frac{2 J_{\rm H}}{3} m_2 ^{x,y} \nonumber \\
2 \Delta^z _{2} &=& {\mathcal U}_2 m_2 ^z + \frac{2J_{\rm H}}{3} m_1 ^z - J_{\rm H} m_1 ^z  \nonumber \\
2 \Delta^{x,y} _{2} &=& {\mathcal U}_2 m_2 ^{x,y} + \frac{2 J_{\rm H}}{3} m_1 ^{x,y} \nonumber \\
2 \Delta^z _{3} &=& {\mathcal U}_3 m_3 ^z + J_{\rm H} m_1 ^z  \nonumber \\
2 \Delta^{x,y} _{3} &=& {\mathcal U}_3 m_3 ^{x,y} 
\label{selfcon}
\end{eqnarray}
in terms of the staggered pseudo-spin magnetizations ${\bf m}_l$=$m_l ^x,m_l ^y,m_l ^z$ for the three pseudo-orbitals $l=1, 2, 3$. In practice, it is easier to choose set of ${\bf \Delta}_{l=1,2,3}$ and self-consistently determine the Hubbard-like interaction strengths ${\mathcal U}_{l=1,2,3}$ such that ${\mathcal U}_1 = U - \frac{4}{3} J_{\rm H}$ and ${\mathcal U}_2 = {\mathcal U}_3 =  U - J_{\rm H}$ using Eq. (\ref{selfcon}). The interaction strengths are related by ${\mathcal U}_{l=2,3} = {\mathcal U}_{l=1} + J_{\rm H}/3$.

Transforming back to the three-orbital basis $(yz\sigma,xz\sigma,xy\bar{\sigma})$, the staggered-field contribution for the $l=1$ sector is illustrated below: 
\begin{equation}
\mathcal{H}_{\rm sf} ^{l=1} = \sum_{{\bf k} \sigma s}  s \sigma
\psi_{{\bf k} \sigma s}^{\dagger}
\left [
 \frac{ \Delta_1 ^z}{3}
\begin{pmatrix}
1 & i \sigma & -\sigma \\
-i \sigma & 1 & i \\
-\sigma & -i & 1 \\
\end{pmatrix} \delta_{\sigma \sigma'}
+
\left( \frac{-\Delta_1 ^x + i \Delta_1 ^y}{3} \right)
\begin{pmatrix}
1 & i \sigma & -\sigma \\
-i \sigma & -1 & i \\
-\sigma & -i & -1 \\
\end{pmatrix} \delta_{\bar{\sigma} \sigma'}
 \right] \psi_{{\bf k} \sigma' s} 
\label{stag_field_matrix}
\end{equation} 
which has similar structure as the spin-orbit coupling term. Including the SOC and band terms, the full HF Hamiltonian considered in our band structure and spin fluctuation analysis is given by
$\mathcal {H}_{\rm HF} = \mathcal{H}_{\rm SO} + \mathcal{H}_{\rm band} + \mathcal{H}_{\rm sf}$,
where,
\begin{eqnarray}
\mathcal{H}_{\rm SO} + \mathcal{H}_{\rm band} &=& 
\sum_{{\bf k} \sigma s} \psi_{{\bf k} \sigma s}^{\dagger} \left [ \begin{pmatrix}
{\epsilon_{\bf k} ^{yz}}^\prime & i \sigma\frac{\lambda}{2} & -\sigma\frac{\lambda}{2} \\
- i \sigma\frac{\lambda}{2} & {\epsilon_{\bf k} ^{xz}}^\prime & i\frac{\lambda}{2} \\
-\sigma\frac{\lambda}{2} & - i\frac{\lambda}{2} & {\epsilon_{\bf k} ^{xy}}^\prime \end{pmatrix} \delta_{s s^\prime}
+ 
\begin{pmatrix}
\epsilon_{\bf k} ^{yz} & \epsilon_{\bf k} ^{yz|xz} & 0 \\
-\epsilon_{\bf k} ^{yz|xz} & \epsilon_{\bf k} ^{xz} & 0 \\
0 & 0 & \epsilon_{\bf k} ^{xy} \end{pmatrix} \delta_{\bar{s} s^\prime } \right]
\psi_{{\bf k} \sigma s^\prime} \nonumber \\
\label{three_orb_two_sub}
\end{eqnarray} 
in the composite three-orbital, two-sublattice basis, showing the different hopping terms connecting the same and opposite sublattice(s). 

Corresponding to the hopping terms in the tight-binding model, the various band dispersion terms in Eq. (\ref{three_orb_two_sub}) are given by: 
\begin{eqnarray}
\epsilon_{\bf k} ^{xy} &=& -2t_1(\cos{k_x} + \cos{k_y}) \nonumber \\
{\epsilon_{\bf k} ^{xy}} ^{\prime} &=& - 4t_2\cos{k_x}\cos{k_y} - \> 2t_3(\cos{2{k_x}} + \cos{2{k_y}}) + \mu_{xy}  \nonumber \\
\epsilon_{\bf k} ^{yz} &=& -2t_5\cos{k_x} -2t_4 \cos{k_y} \nonumber \\
\epsilon_{\bf k} ^{xz} &=& -2t_4\cos{k_x} -2t_5 \cos{k_y}  \nonumber \\
\epsilon_{\bf k} ^{yz|xz} &=&  -2t_{m}(\cos{k_x} + \cos{k_y}) . 
\end{eqnarray}
Here $t_1$, $t_2$, $t_3$ are respectively the first, second, and third neighbor hopping terms for the $xy$ orbital, which has energy offset $\mu_{xy}$ from the degenerate $yz/xz$ orbitals induced by the tetragonal splitting. For the $yz$ ($xz$) orbital, $t_4$ and $t_5$ are the NN hopping terms in $y$ $(x)$ and $x$ $(y)$ directions, respectively. Mixing between $xz$ and $yz$ orbitals is represented by the NN hopping term $t_m$. We have taken values of the tight-binding parameters ($t_1$, $t_2$, $t_3$, $t_4$, $t_5$, $t_{\rm m}$, $\mu_{xy}$, $\lambda$) = (1.0, 0.5, 0.25, 1.028, 0.167, 0.2, -0.7, 1.35) in units of $t_1$, where the energy scale $t_1$ = 280 meV. Using above parameters, the calculated electronic band structure shows AFM insulating state and mixing between pseudo-orbital sectors.\cite{watanabe_PRL_2010,iridate1_PRB_2017}

\subsection*{Canted AFM state and $J_{\rm H}$ induced easy-plane anisotropy}

\begin{figure}
\vspace*{0mm}
\hspace*{0mm}
\psfig{figure=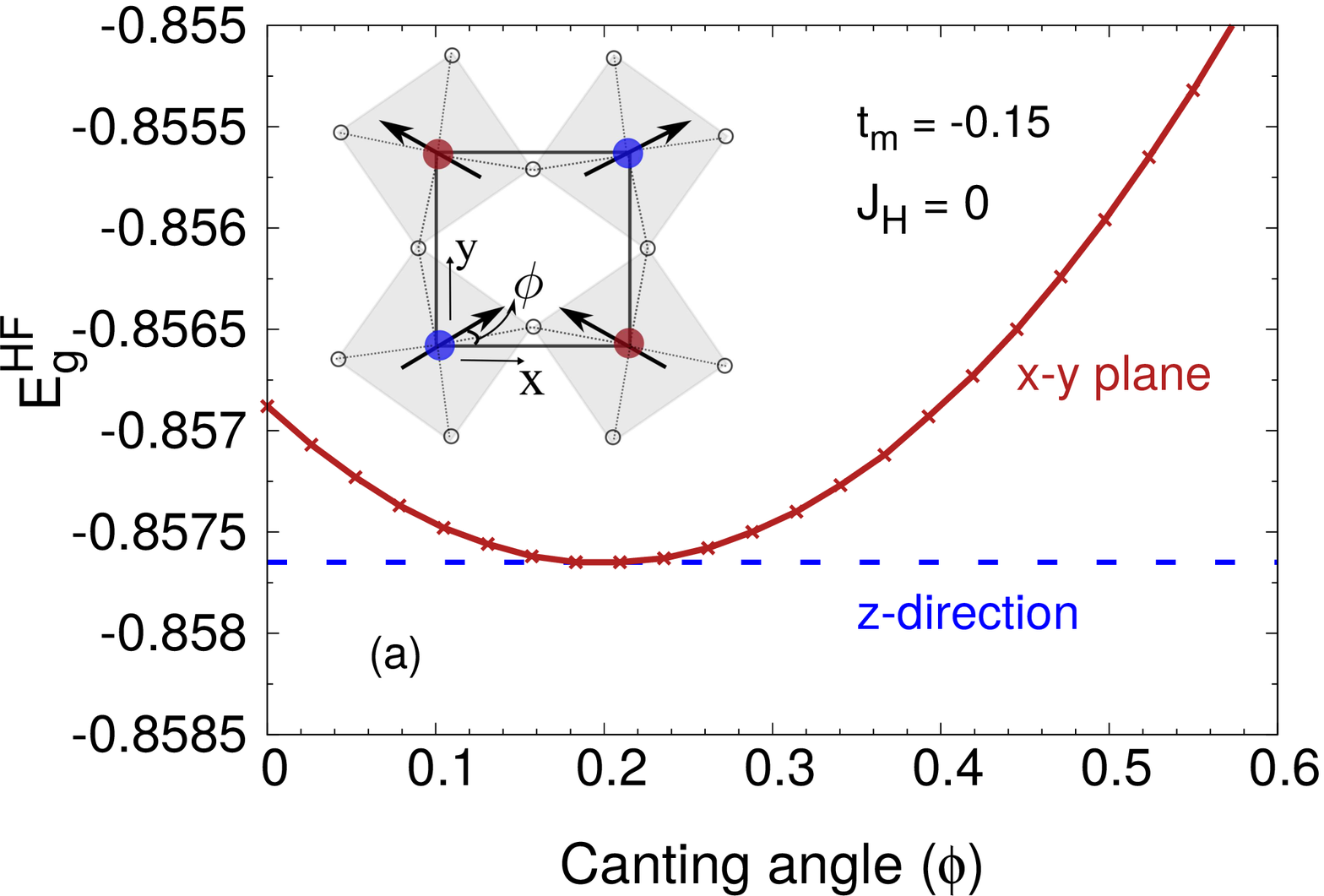,angle=0,width=80mm} 
\psfig{figure=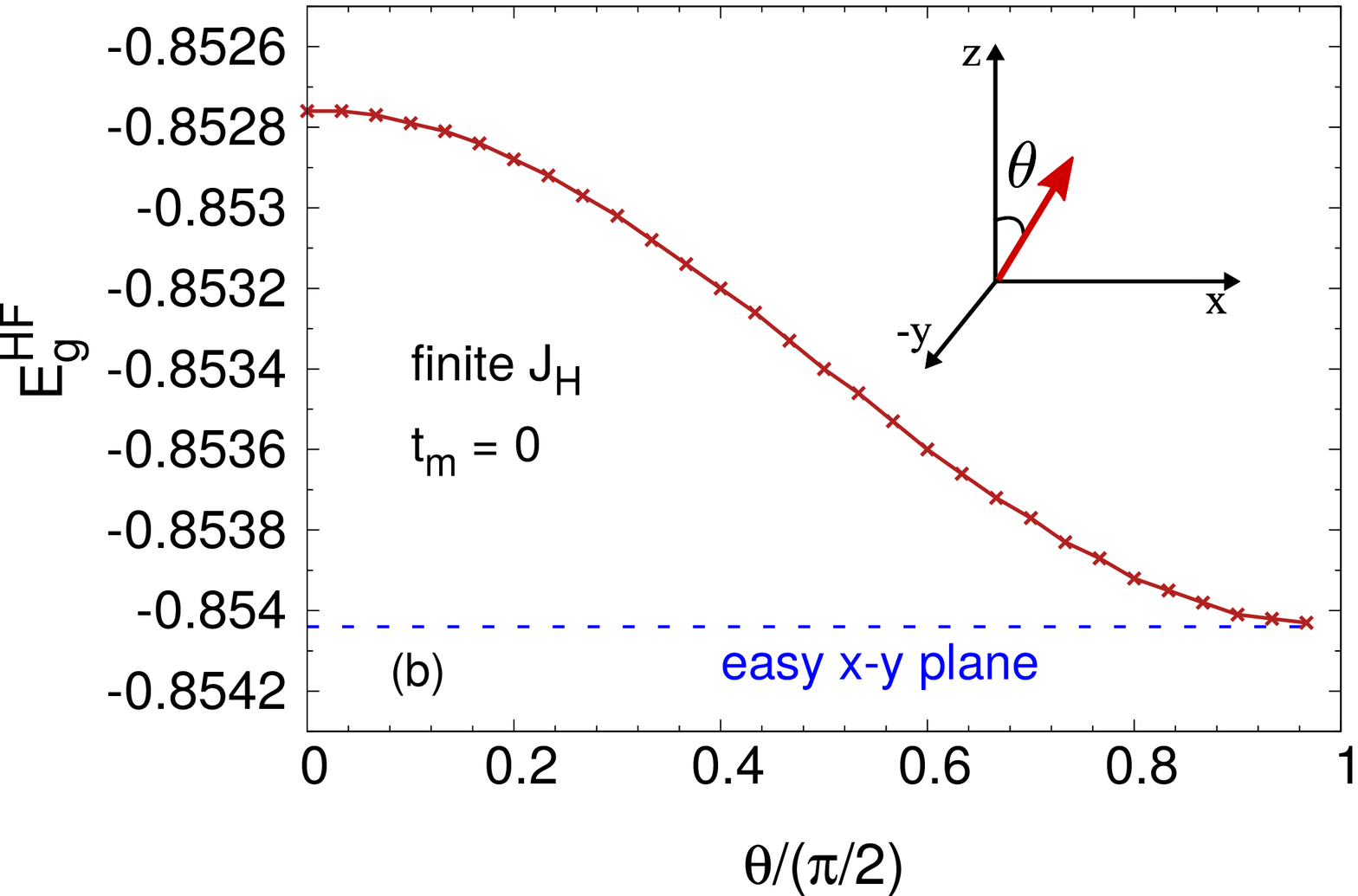,angle=0,width=80mm} 
\vspace{-0mm}
\caption{Variation of AFM state energy per state (a) for $x$-$y$ plane ordering with canting angle $\phi$ including finite $yz-xz$ orbital mixing hopping showing degeneracy at the optimal canting angle with the $z$-ordered AFM state, and (b) with the staggered field polar angle $\theta$ showing easy $x$-$y$ plane anisotropy for finite Hund's coupling.} 
\label{gse}
\end{figure}

The octahedral-rotation-induced orbital mixing hopping term ($t_{\rm m}$) between $yz$ and $xz$ orbitals generates PD ($S_i^z S_j^z$) and DM [$(\vec S_i \times \vec S_j). \hat z$] anisotropic interactions in the strong coupling limit.\cite{iridate1_PRB_2017} However, the AFM-state energy is invariant with respect to change of ordering direction from $z$ axis to $x$-$y$ plane provided spins are canted at the optimal canting angle, thus preserving the gapless Goldstone mode. Fig. \ref{gse}(a) shows the variation of AFM-state energy with canting angle ($\phi$) for ordering in the $x$-$y$ plane. The energy minimum at the optimal canting angle is exactly degenerate with the energy for $z$-direction ordering.

The Hund's-coupling-induced easy-plane magnetic anisotropy is explicitly shown in Fig.\ref{gse}(b) by the variation of AFM-state energy with polar angle $\theta$ corresponding to staggered field orientation in the $x$-$z$ plane, with $\Delta_1 ^z = (\Delta +\Delta_{\rm ani})\cos \theta$ and $\Delta_1 ^x = \Delta \sin \theta$. Here $\Delta$ represents the spin-rotationally-symmetric part [$({\mathcal U}_1 m_1 + \frac{2J_{\rm H}}{3} m_2)/2$] of the staggered field term for $l$=1 orbital. The symmetry-breaking term $\Delta_{\rm ani}$ corresponds to the additional contribution $\frac{1}{2}J_{\rm H} (m_3 ^z - m_2 ^z)$, as seen from Eq. (\ref{selfcon}). Here $\Delta$=0.9, $\Delta_{\rm ani}$=$-0.01$, and the orbital mixing hopping term $t_{\rm m}$ has been set to zero for simplicity. The simplified analysis presented in this subsection, with staggered field only for the $l$=1 orbital, serves to explicitly illustrate the magnetic anisotropy features within our band picture. 

\section{Magnetic anisotropy and gapped spin wave}

In view of the Hund's-coupling-induced easy-plane anisotropy as discussed above, we consider the $x$-ordered AFM state. The spin-wave propagator corresponding to transverse spin fluctuations should therefore yield one gapless mode ($y$ direction) and one gapped mode ($z$ direction). Accordingly, we consider the time-ordered spin-wave propagator: 
\begin{equation}
\chi ({\bf q},\omega) = \int dt \sum_{i} e^{i\omega(t-t^\prime)}
e^{-i{\bf q}.({\bf r}_i - {\bf r}_j)}  
\langle \Psi_0 | T [ S_{i m} ^{\alpha} (t) S_{j n} ^{\beta} (t^\prime) ] | \Psi_0 \rangle
\label{chi}
\end{equation}
involving the transverse $\alpha,\beta=y,z$ components of the pseudo-spin operators $S_{im}^{\alpha}$ and $S_{jn}^{\beta}$ for pseudo orbitals $m$ and $n$ at lattice sites $i$ and $j$. 

In the random phase approximation (RPA), the spin-wave propagator is obtained as:
\begin{equation}
[\chi({\bf q},\omega)] = \frac{[\chi^0({\bf q},\omega)]}
{1 - 2 [\mathcal{U}] [\chi^0({\bf q},\omega)]}
\label{eq:spin_prop}  
\end{equation}
where the bare particle-hole propagator:
\begin{equation}
[\chi^0 ({\bf q},\omega)]_{a b} ^{\alpha \beta} = \frac{1}{4} \sum_{{\bf k}} \left [ 
\frac{ 
\langle \varphi_{\bf k-q} | \tau^\alpha | \varphi_{\bf k} \rangle_a
\langle \varphi_{\bf k} | \tau^\beta | \varphi_{\bf k-q} \rangle_b 
} 
{E^+_{\bf k-q} - E^-_{\bf k} + \omega - i \eta }
+ \frac{
\langle \varphi_{\bf k-q} | \tau^\alpha | \varphi_{\bf k} \rangle_a
\langle \varphi_{\bf k} | \tau^\beta | \varphi_{\bf k-q} \rangle_b 
} 
{E^+_{\bf k} - E^-_{\bf k-q} - \omega - i \eta } \right ]
\end{equation}
was evaluated in the composite spin-orbital-sublattice basis (2 spin components $\alpha,\beta=y,z$ $\otimes$ 3 pseudo orbitals $m=1,2,3$ $\otimes$ 2 sublattices $s,s'=$ A,B) by integrating out the fermions in the $(\pi,\pi)$ ordered state. Here $E_{\bf k}$ and $\varphi_{\bf k}$ are the eigenvalues and eigenvectors of the Hamiltonian matrix in the pseudo-orbital basis, the indices $a,b=1,6$ correspond to the orbital-sublattice subspace, and the superscript $+(-)$ refers to particle (hole) energies above (below) the Fermi energy. The amplitudes $\varphi^m _{{\bf k}\tau}$ were obtained by projecting the ${\bf k}$ states in the three-orbital basis on to the pseudo-orbital basis states $|m,\tau = \uparrow,\downarrow \rangle$  corresponding to the $J=1/2$ and $3/2$ sector states, as given below:
\begin{eqnarray}
\varphi_{{\bf k}\uparrow}^1 = \frac{1}{\sqrt{3}} \left( \phi^{yz}_{{\bf k}\downarrow} - i\phi^{xz}_{{\bf k}\downarrow} + \phi^{xy}_{{\bf k}\uparrow}\right) \;\;\;\;\;\; & & 
\varphi_{{\bf k}\downarrow}^1 = \frac{1}{\sqrt{3}} \left( \phi^{yz}_{{\bf k}\uparrow} + i \phi^{xz}_{{\bf k}\uparrow} - \phi^{xy}_{{\bf k}\downarrow}\right) \nonumber \\ 
\varphi_{{\bf k}\uparrow}^2 = \frac{1}{\sqrt{6}} \left( \phi^{yz}_{{\bf k}\downarrow} - i\phi^{xz}_{{\bf k}\downarrow} - 2 \phi^{xy}_{{\bf k}\uparrow}\right)  \;\;\;\; & & 
\varphi_{{\bf k}\downarrow}^2 = \frac{1}{\sqrt{6}} \left( \phi^{yz}_{{\bf k}\uparrow} + i \phi^{xz}_{{\bf k}\uparrow} + 2 \phi^{xy}_{{\bf k}\downarrow}\right) \nonumber \\
\varphi_{{\bf k}\uparrow}^3 = \frac{1}{\sqrt{2}} \left( \phi^{yz}_{{\bf k}\downarrow} + i\phi^{xz}_{{\bf k}\downarrow} \right) \;\;\;\;\;\;\;\;\;\;\;\;\;\;\; & & 
\varphi_{{\bf k}\downarrow}^3 = \frac{1}{\sqrt{2}} \left( \phi^{yz}_{{\bf k}\uparrow} - i \phi^{xz}_{{\bf k}\uparrow} \right)
\label{proj_ampl}
\end{eqnarray}
in terms of the amplitudes $\phi^{\mu}_{{\bf k}\sigma}$ in the three-orbital basis $(\mu = yz,xz,xy)$. 

The rotationally invariant Hubbard- and Hund's coupling-like terms having the form $S_{im}^\alpha S_{in} ^\beta \delta_{\alpha\beta}$ are diagonal in spin components ($\alpha = \beta$). The on-site Coulomb interaction terms are also diagonal in the sublattice basis ($s=s'$). The interaction matrix $[\mathcal{U}]$ in Eq. (\ref{eq:spin_prop}) is therefore obtained as:
\begin{equation}
[\mathcal{U}] = \begin{pmatrix} \mathcal{U}_1 & \frac{2}{3} J_{\rm H} & 0 \\
\frac{2}{3} J_{\rm H} & \mathcal{U}_2 & 0 \\ 0 & 0 & \mathcal{U}_3 
\end{pmatrix} \delta_{\alpha\beta} \delta_{ss'} + 
\begin{pmatrix} 0 & -J_{\rm H} & J_{\rm H}  \\
-J_{\rm H} & 0 & 0 \\ J_{\rm H} & 0 & 0 
\end{pmatrix} \delta_{\alpha z} \delta_{\beta z} \delta_{ss'} 
\end{equation}
in the pseudo-orbital basis. While the first interaction term above preserves spin rotation symmetry, the second interaction term (corresponding to the $S_{im}^z S_{in}^z$ terms in Eq. \ref{h_int}) breaks rotation symmetry and is responsible for easy $x$-$y$ plane anisotropy. The spin wave energies are calculated from the poles of Eq. \ref{eq:spin_prop}. The $12 \times 12$  $[\chi^0 ({\bf q},\omega)]$ matrix was evaluated by performing the $\bf k$ sum over the 2D Brillouin zone divided into a 300 $\times$ 300 mesh.

\begin{figure}
\vspace*{0mm}
\hspace*{0mm}
\psfig{figure=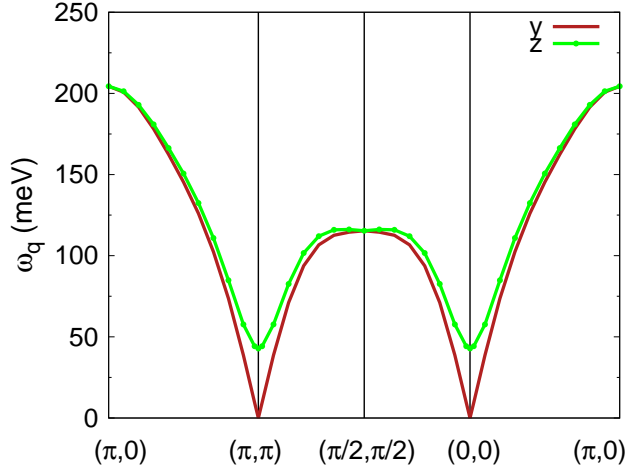,angle=0,width=90mm}
\vspace{0mm}
\caption{The calculated spin-wave dispersion in the three-orbital model with staggered field in the $x$ direction. The easy $x$-$y$ plane anisotropy arising from Hund's coupling results in one gapless mode and one gapped mode corresponding to transverse fluctuations in the $y$ and $z$ directions, respectively.}
\label{sw}
\end{figure}

\section{Spin-wave Dispersion}
The calculated spin-wave energies in the $x$-ordered AFM state are shown in Fig. \ref{sw}. Here we have taken staggered field values $\Delta_{l=1,2,3}^x= (0.92,0.08,-0.06)$ in units of $t_1$, which ensures self-consistency for all three orbitals, with the given relations ${\mathcal U}_2$=${\mathcal U}_3$=${\mathcal U}_1$+$J_{\rm H}/3$. Using the calculated sublattice magnetization values $m_{l=1,2,3}^x$=(0.65,0.005,-0.038), we obtain ${\mathcal U}_{l=1,2,3}$=(0.80,0.83,0.83) eV, which finally yields $U$=$\mathcal U_1$+$\frac{4}{3}J_{\rm H}$=0.93 eV for $J_{\rm H}$=0.1 eV. 

The spin-wave dispersion clearly shows the Goldstone mode and the gapped mode, corresponding to transverse spin fluctuations in the $y$ and $z$ directions, respectively. The easy $x$-$y$ plane anisotropy arising from Hund's coupling results in energy gap $\approx$40 meV for the out-of-plane ($z$) mode. The two modes are degenerate at $(\pi,0)$ and $(\pi/2,\pi/2)$. The excitation energy at $(\pi,0)$ is approximately twice that at $(\pi/2,\pi/2)$, and the strong zone-boundary dispersion in this iridate compound was ascribed to finite-$U$ and finite-SOC effects.\cite{iridate1_PRB_2017} The calculated spin-wave dispersion and energy gap are in very good agreement with RIXS measurements.\cite{kim1_PRL_2012,kim_NATCOMM_2014,pincini_PRB_2017,porras_arxiv_2018}


The electron fillings in the different pseudo orbitals are obtained as $n_{l=1,2,3} \approx (1.064,1.99,1.946)$. Finite mixing between the $J$=1/2 and 3/2 sectors is reflected in the small deviations from ideal fillings and also in the very small magnetic moment values for $l=2,3$ as given above, which play a crucial role in the expression of magnetic anisotropy and spin-wave gap in view of the anisotropic interaction terms in Eq. (\ref{h_int}) involving the Hund's coupling $J_{\rm H}$. The values $\lambda$=0.38 eV, $U$=0.93 eV, and $J_{\rm H}$=0.1 eV taken above lie well within the estimated parameter range for $\rm Sr_2 Ir O_4$.\cite{zhou_PRX_2017,igarashi_PRB_2014}

\begin{figure}
\vspace*{0mm}
\hspace*{0mm}
\psfig{figure=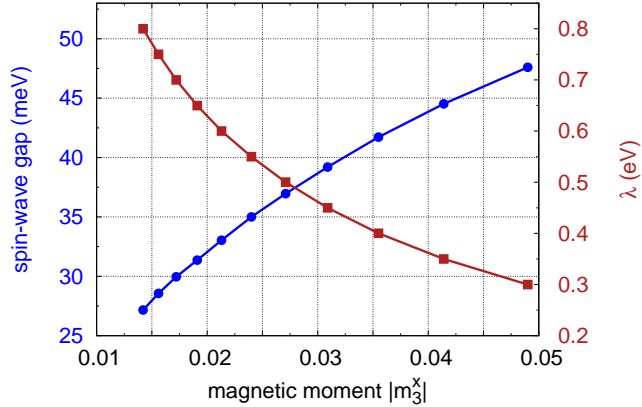,angle=0,width=90mm}
\vspace{0mm}
\caption{Variation of spin-wave gap with magnetic moment $|m_3^x|$ in the $l$=3 orbital (blue curve). The magnitude of $|m_3^x|$ decreases with SOC strength (red curve) due to suppression of mixing between $J$=1/2 and 3/2 sectors. Here $(U,J_{\rm H})$=(0.93 eV, 0.1 eV).}
\label{fig3b}
\end{figure}

We have investigated the crucial role of the small $J$=3/2-sector magnetic moment by studying the variation of the spin-wave gap with SOC strength which effectively controls the mixing between $J$=1/2 and 3/2 sectors. Fig. \ref{fig3b} shows that the spin-wave gap sharply increases with magnetic moment $|m_3^x|$ in the $l$=3 orbital (the dominant moment), indicating a finite-SOC effect on the experimentally observed out-of-plane spin-wave gap in $\rm Sr_2 Ir O_4$. The opposite sign of the magnetic moment $m_3^x$ as compared to $m_1^x$ (due to spin-orbital entanglement) plays a vital role in the easy-plane anisotropy.  

\section{Renormalized spin-orbit gap}

As another application of the transformation described in Sec. III, we now consider the relative energy shift between the $J$=1/2 and 3/2 states arising from the density interaction terms in Eq. (\ref{h_int}). This relative shift effectively renormalizes the spin-orbit gap and plays an important role in determining the energy scale of the spin-orbit exciton, as discussed in the next section. Corresponding to the total density condensate $\langle n_{l\uparrow} + n_{l\downarrow} \rangle$ in the HF approximation of the density interaction terms, the spin-independent self-energy contributions for the three orbitals are obtained as: 
\begin{eqnarray}
\Sigma_{\rm dens}^{l=1} &=& U \left \langle \frac{1}{2}n_1 + n_2 + n_3 \right \rangle - J_{\rm H} \left \langle \frac{2}{3}n_1 + \frac{13}{6} n_2 + \frac{13}{6} n_3 \right \rangle  \nonumber \\
\Sigma_{\rm dens} ^{l=2} &=& U \left \langle n_1 + \frac{1}{2}n_2 + n_3 \right \rangle - J_{\rm H} \left \langle \frac{13}{6}n_1 + \frac{1}{2} n_2 + \frac{7}{3} n_3 \right \rangle \nonumber \\
\Sigma_{\rm dens} ^{l=3} &=& U \left \langle n_1 + n_2 + \frac{1}{2}n_3 \right \rangle - J_{\rm H} \left \langle \frac{13}{6}n_1 + \frac{7}{3} n_2 + \frac{1}{2} n_3 \right \rangle 
\end{eqnarray}
The formally unequal contributions will result in relative energy shifts between the three orbitals depending on the electron filling. With $\langle n_1 \rangle$=1 and $\langle n_2 \rangle$=$\langle n_3 \rangle$=2 for the $d^5$ system having nominally half-filled and filled orbitals, the relative energy shift: 
\begin{equation}
\Delta_{\rm dens} = \Sigma_{\rm dens}^{l=1} - \Sigma_{\rm dens}^{l=2,3} = \frac{U- 3 J_{\rm H}}{2}
\end{equation}
between $l$=1 and (degenerate) $l$=2,3 orbitals. 

For $U > 3J_{\rm H}$, the relative energy shift enhances the energy gap between $J$=1/2 and $3/2$ sectors, effectively resulting in a correlation-induced renormalization of the spin-orbit gap and the spin-orbit coupling. For $d^4$ systems with nominally $\langle n_1 \rangle$=0, the relative energy shift increases to $U-3J_{\rm H}$. This enhancement of the spin-orbit gap renormalization is seen in recent DFT study of the hexagonal iridates $\rm Sr_3 Li Ir O_6$ and $\rm Sr_4 Ir O_6$ with Ir$^{5+}$ ($5d^4$) and Ir$^{4+}$ ($5d^5$) ions, respectively.\cite{ming_PRB_2018} 

The SOC strength is renormalized as $\tilde{\lambda} = \lambda + 2\Delta_{\rm dens}/3$ by the relative energy shift. With $\Delta_{\rm dens} = (U-3J_{\rm H})/2 \approx$ 0.3 eV for the parameter values considered earlier, we obtain $\tilde{\lambda} \approx$ 0.6 eV, which is in agreement with the correlation-enhanced SOC strength obtained in a recent DFT study of $\rm Sr_2 Ir O_4$.\cite{zhou_PRX_2017} The SOC renormalization also improves the comparison of spin-wave dispersion with experiment near $(\pi/2,\pi/2)$ as shown in Fig. 5(a). With the bare SOC strength, the collective spin-wave mode is squeezed by the particle-hole excitation, as seen in Fig. 5(b). The renormalized spin-orbit gap increases the particle-hole excitation energy and thereby removes the flattening. By effectively suppressing the mixing between $J$=1/2 and 3/2 sectors, the SOC renormalization also strengthens the AFM state. 


\begin{figure}
\vspace*{0mm}
\hspace*{0mm}
\psfig{figure=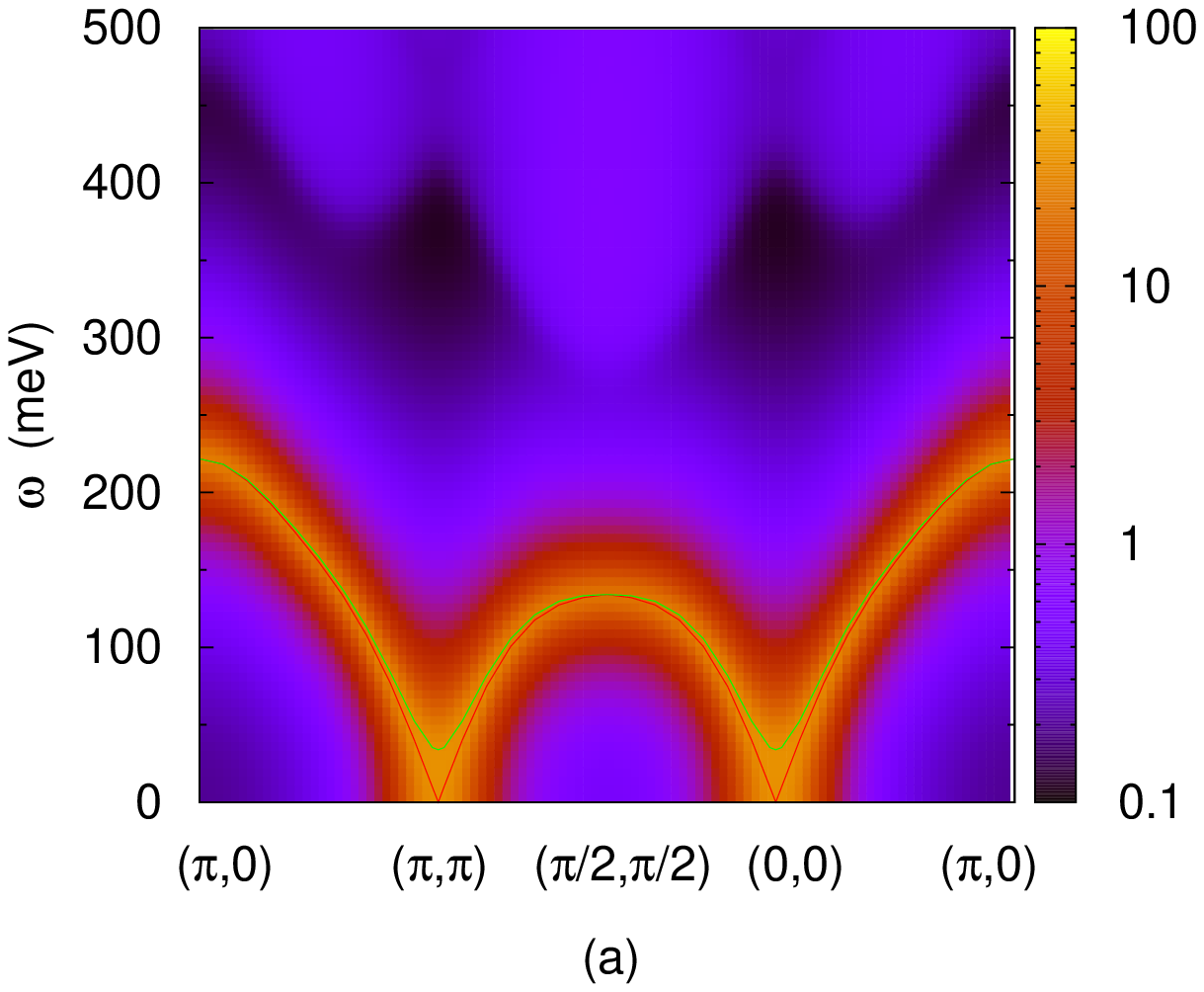,angle=0,width=80mm}
\psfig{figure=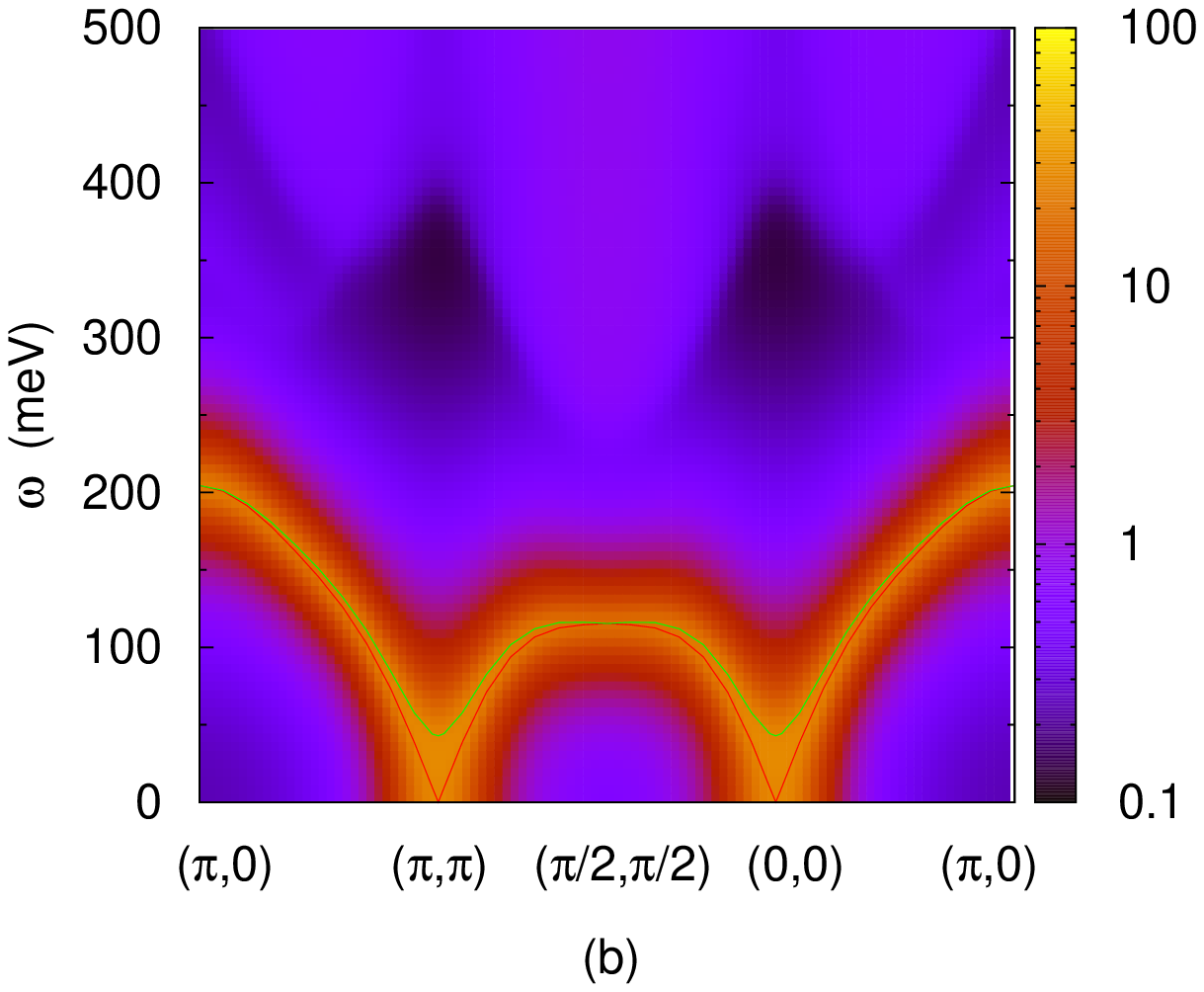,angle=0,width=80mm}
\vspace*{-5mm}
\caption{(a) Calculated spin-wave spectral function with (a) the renormalized SOC and (b) the bare SOC, showing the squeezing of the spin-wave mode by the particle-hole excitations near $(\pi/2,\pi/2)$.}
\label{fig5}
\end{figure}

\section{Spin-Orbit Exciton}

The low-energy collective (spin-wave ) modes investigated in Secs. V and VI essentially involve intra-orbital spin-flip excitations within the magnetically active $J$=1/2 sector. In this section, we will investigate inter-orbital, spin-flip, particle-hole excitations across the spin-orbit gap between the nominally filled $J$=3/2 sector and the half-filled $J$=1/2 sector. For the $z$-ordered AFM state, we consider the composite pseudo-spin-orbital fluctuation propagator:
\begin{equation}
\chi^{-+} _{\rm so} ({\bf q},\omega) = \int dt \sum_{i} e^{i\omega(t-t^\prime)}
e^{-i{\bf q}.({\bf r}_i - {\bf r}_j)}  
\langle \Psi_0 | T [ S_{i,m,n} ^- (t) S_{j,m,n} ^+ (t^\prime) ] | \Psi_0 \rangle
\label{soe_prop}
\end{equation}
involving the inter-orbital spin-lowering and -raising operators $S_{i,m,n}^-$=$a_{in\downarrow} ^\dagger a_{im\uparrow}$ and $S_{j,m,n}^+$=$a_{jm\uparrow} ^\dagger a_{jn\downarrow}$ at lattice sites $i$ and $j$, describing the propagation of a spin-flip particle-hole excitation between different pseudo orbitals $m$ and $n$. Although the most general propagator would involve $S_{i,m,n}^-$ and $S_{j,m',n'}^+$, the above simplified propagator is a good approximation in view of the orbital restrictions on the particle-hole states as discussed below. Also, we have considered the $z$-ordered AFM state as the weak easy-plane anisotropy has negligible effect on the spin-orbit exciton. 

In the ladder-sum approximation, the spin-orbital propagator is obtained as:
\begin{equation}
[\chi^{-+} _{\rm so} ({\bf q},\omega)] = \frac{[\chi^0 _{\rm so} ({\bf q},\omega)]}
{1 - {\mathcal U} [\chi^0 _{\rm so} ({\bf q},\omega)]}
\label{soe_RPA}  
\end{equation}
where the relevant interactions ${\mathcal U} = \mathcal{U}_{mn}^{\tau \overline{\tau}}$ for the spin-flip particle-hole pair are given in Eq. (\ref{cal_u_eqns}), and the bare particle-hole propagator:
\begin{equation}
[\chi^0 _{\rm so} ({\bf q},\omega)]_{s s'} ^{mn} = \sum_{{\bf k}} \left [ 
\frac{ 
\langle \varphi_{\bf k-q} ^n | \tau^- | \varphi_{\bf k} ^m \rangle_s
\langle \varphi_{\bf k} ^m   | \tau^+ | \varphi_{\bf k-q} ^n \rangle_{s'} 
} 
{E^+_{\bf k-q} - E^-_{\bf k} + \omega - i \eta }
+ \frac{
\langle \varphi_{\bf k-q} ^n | \tau^- | \varphi_{\bf k} ^m \rangle_s
\langle \varphi_{\bf k} ^m   | \tau^+ | \varphi_{\bf k-q} ^n \rangle_{s'} 
} 
{E^+_{\bf k} - E^-_{\bf k-q} - \omega - i \eta } \right ]
\label{chi0_so}
\end{equation}
was evaluated using the projected amplitudes given in Eq. \ref{proj_ampl}. The ladder-sum approximation with repeated (attractive) interactions represents resonant scattering of the particle-hole pair, resulting in a resonant state split-off from the particle-hole continuum, which we identify as the spin-orbit exciton mode. 

The dominant contribution to the bare particle-hole propagator above will correspond to particle $(+)$ states in the nominally half-filled pseudo-orbital $m$=1 ($J$=1/2 sector) and hole $(-)$ states in the nominally filled pseudo-orbitals $n$=2,3 ($J$=3/2 sector). Due to these restrictions, the bare propagator essentially becomes diagonal in the composite particle-hole orbital basis ($m'$=$m$,$n'$=$n$), which justifies the simplified propagator considered above. In order to focus exclusively on the high-energy spin-orbit exciton mode, particle-hole excitations within the $J$=1/2 sector (which yield the low-energy spin-wave  mode) have been excluded. 

\begin{figure}
\vspace*{0mm}
\hspace*{0mm}
\psfig{figure=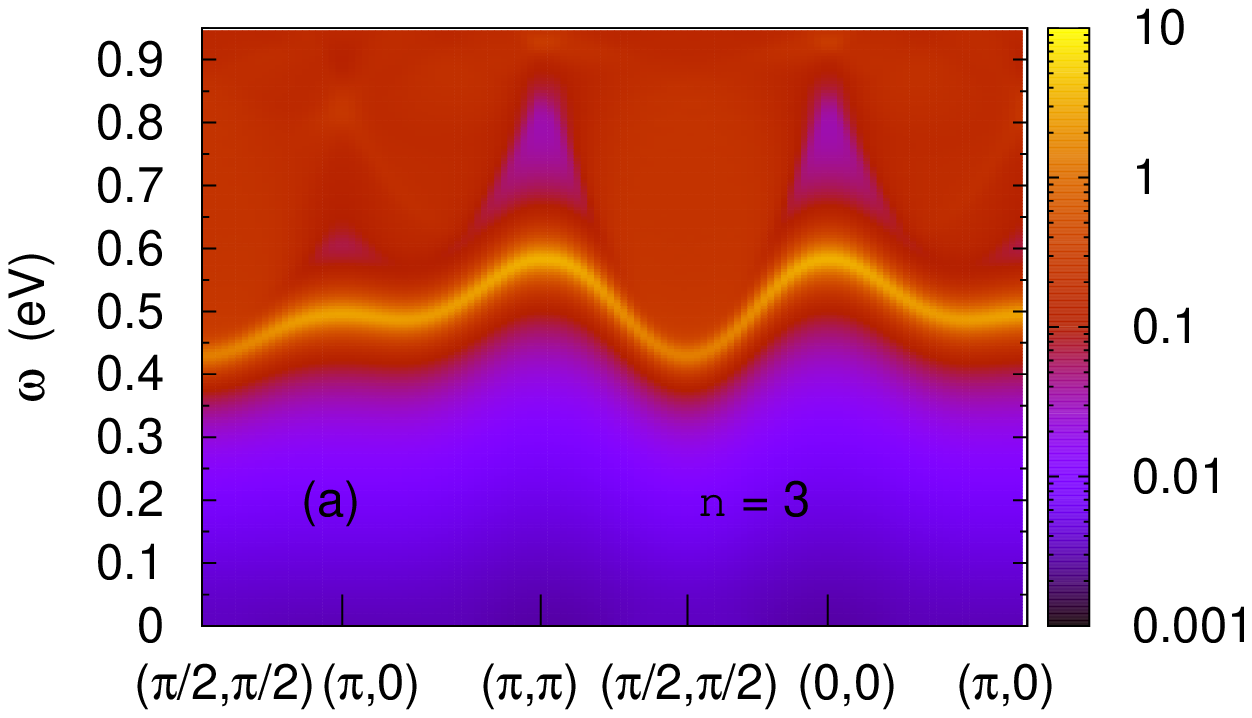,angle=0,width=80mm}
\psfig{figure=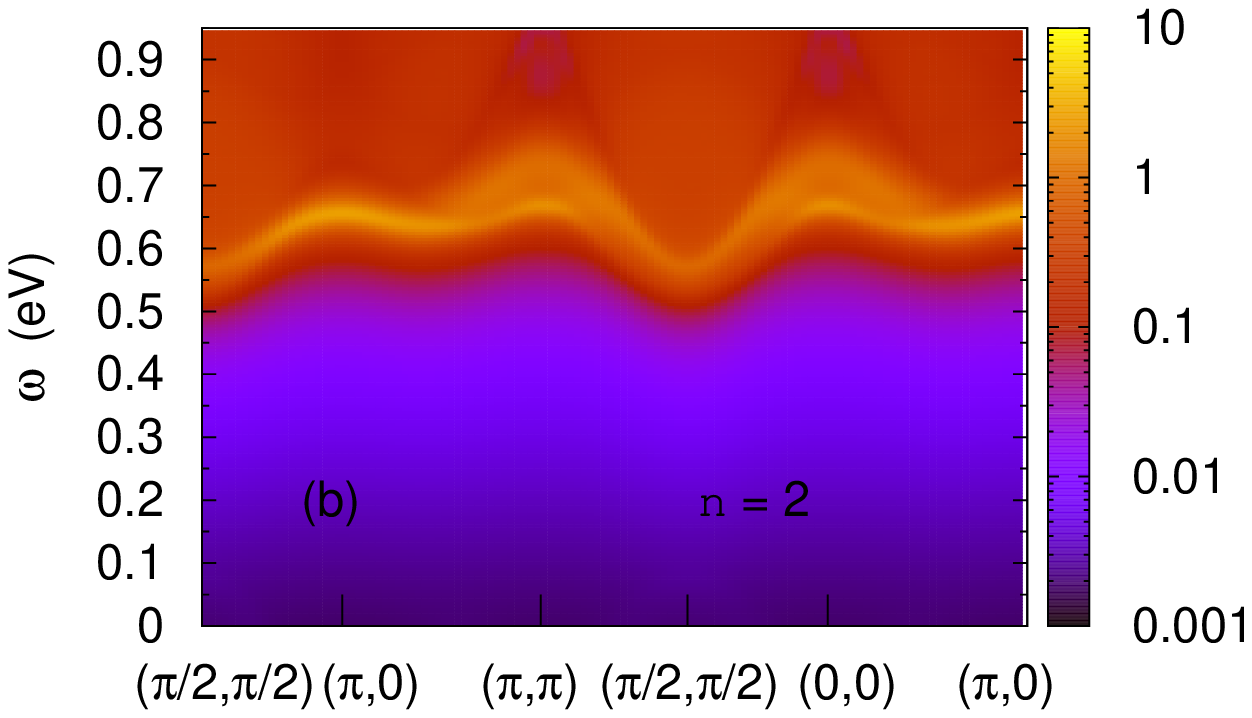,angle=0,width=80mm}
\vspace{-10mm}
\caption{The spin-orbit exciton spectral function $A_{\bf q}(\omega)$ for the two cases: (a) $n$=3 and (b) $n$=2, showing well defined dispersive modes near the lower edge of the continuum. The exciton represents collective spin-orbital excitations across the renormalized spin-orbit gap.}
\label{spectral}
\end{figure}

Fig. \ref{spectral} shows the spin-orbit exciton spectral function:
\begin{equation}
A_{\bf q}(\omega) = \frac{1}{\pi} {\rm Im \; Tr} \left [ \chi^{-+} _{\rm so} ({\bf q},\omega) \right] 
\end{equation}
as an intensity plot for $\bf q$ along the high symmetry directions of the BZ. For clarity, we have considered here the particle-hole propagator for $m$=1 and $n$=3,2 separately in Eq. (\ref{chi0_so}), for which the relevant interaction terms are: $\mathcal{U}_{13}^{\tau \overline{\tau}}$=$U$-$5J_{\rm H}/3$ and $\mathcal{U}_{12}^{\tau \overline{\tau}}$=$U$-$7J_{\rm H}/3$. Here, we have taken $U$=0.93 eV and $J_{\rm H}$=0.1 eV as in Sec. VI, the three-orbital model parameters are same as in Sec. IV, and the renormalized spin-orbit gap has been incorporated. 

The spin-orbit exciton spectral function in Fig. \ref{spectral}(a) clearly shows a well defined propagating mode near the lower edge of the continuum with significantly higher intensity compared to the continuum background. With increasing interaction strength, this mode progressively shifts to lower energy further away from the continuum, and becomes more prominent in intensity, confirming its distinct identity from the continuum background. 

Fig. \ref{spectral}(b) shows a similar exciton mode for the other case ($m$=1,$n$=2), with slightly higher energy and reduced dispersion as well as significant damping. The relatively reduced interaction strength $\mathcal{U}_{12}^{\tau \overline{\tau}}$ for this mode accounts for the slightly higher energy. The calculated dispersion and energy scale of the two spin-orbit exciton modes are in excellent agreement with the two exciton modes reported in RIXS investigations\cite{kim1_PRL_2012,pincini_PRB_2017} of $\rm Sr_2 Ir O_4$ as well as previous theoretical studies.\cite{igarashi_PRB_2014} 


\section{Conclusions}

Transformation of the various Coulomb interaction terms to the pseudo-orbital basis formed by the $J$=1/2 and 3/2 states was shown to provide a versatile tool for investigating magnetic anisotropy effects as well as the spin-orbit exciton modes in the strongly spin-orbit coupled compound $\rm Sr_2 Ir O_4$. Explicitly pseudo-spin-symmetry-breaking terms were obtained (dominantly $\sim J_{\rm H} S_{1}^z S_{3}^z$), resulting in easy $x$-$y$ plane anisotropy and gap for the out-of-plane spin-wave mode, reflecting the importance of mixing with the $J$=3/2 sector in determining the magnetic properties of this compound. 

Well-defined propagating spin-orbit exciton modes were obtained representing collective modes of inter-orbital, spin-flip, particle-hole excitations, with both dispersion and energy scale in excellent  agreement with RIXS studies. The relevant interaction terms for the two exciton modes as well as the renormalized spin-orbit gap, which play an important role in the spin-orbit exciton energy scale, were obtained from the transformation, suggesting wider applicability of the general formalism presented here to other spin-orbit coupled systems.


\begin{thebibliography}{99}
\bibitem{krempa_AR_2014} W. Witczak-Krempa, G. Chen, Y. B. Kim, and L. Balents, Annu. Rev. Condens. Matter Phys. {\bf 5}, 57-82 (2014).
\bibitem{rau_AR_2016} J. G. Rau, E. Kin-Ho Lee, and H.-Y. Kee, Annu. Rev. Condens. Matter Phys. {\bf 7}, 195-221 (2016).
\bibitem{bertinshaw_AR_2018} J. Bertinshaw, Y. K. Kim, G. Khaliullin, and B. J. Kim, Annu. Rev. Condens. Matter Phys. (in press).
\bibitem{senthil_PRL_2011} F. Wang and T. Senthil, Phys. Rev. Lett. {\bf 106}, 136402 (2011).
\bibitem{kim3_SC_2014} Y. K. Kim, O. Krupin, J. D. Denlinger, A. Bostwick, E. Rotenberg, Q. Zhao, J. F. Mitchell, J. W. Allen, and B. J. Kim, Science {\bf 345}, 187–190 (2014). 
\bibitem{torre_PRL_2015} A. de la Torre, S. McKeown Walker, F. Y. Bruno, S. Ricc\'o, Z. Wang, I. Gutierrez Lezama, G. Scheerer, G. Giriat, D. Jaccard, C. Berthod, T. K. Kim, M. Hoesch, E. C. Hunter, R. S. Perry, A. Tamai, and F. Baumberger, Phys. Rev. Lett. {\bf 115}, 176402 (2015).
\bibitem{kim4_NAT_2016} Y. K. Kim, N. H. Sung, J. D. Denlinger, and B. J. Kim, Nature Physics {\bf 12}, 37–41 (2016).
\bibitem{gretarsson_PRL_2016} H. Gretarsson, N. Sung, J. Porras, J. Bertinshaw, C. Dietl, Jan A. N. Bruin, A. F. Bangura, Y. K. Kim, R. Dinnebier, J. Kim, A. Al-Zein, M. Moretti Sala, M. Krisch, M. Le Tacon, B. Keimer, and B.  J. Kim, Phys. Rev. Lett. {\bf 117}, 107001 (2016).
\bibitem{chen_NATCOM_2018} X. Chen, J. L. Schmehr, Z. Islam, Z. Porter, E. Zoghlin, K. Finkelstein, J. P. C. Ruff, and S. D. Wilson, Nat. Commun. {\bf 9}, 103 (2018).
\bibitem{bhowal_JPCM_2018}  S. Bhowal, J. M. Kurdestany and S. Satpathy, J. Phys.: Condens. Matter {\bf 30} 235601 (2018).  
\bibitem{kim1_PRL_2012} J. Kim, D. Casa, M. H. Upton, T. Gog, Y.-J. Kim, J. F. Mitchell, M. van Veenendaal, M. Daghofer, J. van den Brink, G. Khaliullin, and B. J. Kim, Phys. Rev. Lett. {\bf 108}, 177003 (2012).
\bibitem{liu_PRB_2016} X. Liu, M. Dean, Z. Meng, M. Upton, T. Qi, T. Gog, Y. Cao, J. Lin, D. Meyers, H. Ding, G. Cao, and J. P. Hill, Phys. Rev. B {\bf 93}, 241102 (2016).
\bibitem{pincini_PRB_2017} D. Pincini, J. G. Vale, C. Donnerer, A. de la Torre, E. C. Hunter, R. Perry, M. Moretti Sala, F. Baumberger, and D. F. McMorrow, Phys. Rev. B {\bf 96}, 075162 (2017).
\bibitem{porras_arxiv_2018} J. Porras, J. Bertinshaw, H. Liu, G. Khaliullin, N. H. Sung, J.-W. Kim, S. Francoual, P. Steffens, G. Deng, M. Moretti Sala, A. Effimenko, A. Said, D. Casa, X. Huang, T. Gog, J. Kim, B. Keimer, and B. J. Kim, arXiv:1808.06920 (2018).
\bibitem{kim_NATCOMM_2014} J. Kim, M. Daghofer, A. H. Said, T. Gog, J. van den Brink, G. Khaliullin, and B. J. Kim, Nat. Commun. {\bf 5}, 4453 (2014).
\bibitem{lu_PRB_2018} X. Lu, P. Olalde-Velasco, Y. Huang, V. Bisogni, J. Pelliciari, S. Fatale, M. Dantz, J. G. Vale, E. C. Hunter, J. Chang, V. N. Strocov, R. S. Perry, M. Grioni, D. F. McMorrow, H. M. R\o{}nnow, and T. Schmitt, Phys. Rev. B {\bf 97}, 041102(R) (2018).
\bibitem{kim2_PRL_2012} B. H. Kim, G. Khaliullin, and B. I. Min, Phys. Rev. Lett. {\bf 109}, 167205 (2012).
\bibitem{igarashi_PRB_2014} J-i Igarashi and T. Nagao, Phys. Rev. B {\bf 90}, 064402 (2014).
\bibitem{khaliullin_PRL_2013} G. Khaliullin, Phys. Rev. Lett. {\bf 111}, 197201 (2013).
\bibitem{sato_PRB_2015} T. Sato, T. Shirakawa, and S. Yunoki, Phys. Rev. B {\bf 91}, 125122 (2015).
\bibitem{valenti_PRL_2017} A. J. Kim, H. O. Jeschke, P. Werner, and R. Valent\'{\i}, Phys. Rev. Lett. {\bf 118}, 086401 (2017).
\bibitem{kaushal_arxiv_2019} N. Kaushal, A. Nocera, G. Alvarez, A. Moreo, and E. Dagotto, arXiv:1901.05578 (2019).
\bibitem{jackeli_PRL_2009} G. Jackeli and G. Khaliullin, Phys. Rev. Lett. {\bf 102}, 017205 (2009).
\bibitem{iridate1_PRB_2017} S. Mohapatra, J. van den Brink, and A. Singh, Phys. Rev. B {\bf 95}, 094435 (2017).
\bibitem{igarashi_PRB_2013} J-i Igarashi and T. Nagao, Phys. Rev. B {\bf 88}, 104406 (2013).
\bibitem{perkins_PRB_2014} N. B. Perkins, Y. Sizyuk, and P. W\"olfle, Phys. Rev. B {\bf 89}, 035143 (2014).
\bibitem{vale_PRB_2015} J. G. Vale, S. Boseggia, H. C. Walker, R. Springell, Z. Feng, E. C. Hunter, R. S. Perry, D. Prabhakaran, A. T. Boothroyd, S. P. Collins, H. M. R\o{}nnow, and D. F. McMorrow, Phys. Rev. B {\bf 92}, 020406(R) (2015).
\bibitem{liu_arxiv_2018} H. Liu and G. Khaliullin, Phys. Rev. Lett. {\bf 122}, 057203 (2019).
\bibitem{igarashi_JPSJ_2014} J-i Igarashi and T. Nagao, J. Phys. Soc. Jpn. {\bf 83}, 053709 (2014).
\bibitem{martins_PRL_2011} C. Martins, M. Aichhorn, L. Vaugier, and S. Biermann, Phys. Rev. Lett. {\bf 107}, 266404 (2011).
\bibitem{arita_PRL_2012} R. Arita, J. Kune\ifmmode \check{s}\else \v{s}\fi{}, A. V. Kozhevnikov, A. G. Eguiluz, and M. Imada, Phys. Rev. Lett. {\bf 108}, 086403 (2012).
\bibitem{zhang_PRL_2013} H. Zhang, K. Haule, and D. Vanderbilt, Phys. Rev. Lett. {\bf 111}, 246402 (2013).
\bibitem{watanabe_PRL_2010} H. Watanabe, T. Shirakawa, and S. Yunoki, Phys. Rev. Lett. {\bf 105}, 216410 (2010).
\bibitem{carter_PRB_2013} J.-M. Carter and H.-Y. Kee, Phys. Rev. B {\bf 87}, 014433 (2013).
\bibitem{zhou_PRX_2017} S. Zhou, K. Jiang, H. Chen, and Z. Wang, Phys. Rev. X {\bf 7}, 041018 (2017).
\bibitem{martins_JPCM_2017} C. Martins, M. Aichhorn, and S. Biermann, J. Phys.: Condens. Matter {\bf 29}, 263001 (2017).
\bibitem{ming_PRB_2018} X. Ming, X. Wan, C. Autieri, J. Wen, and X. Zheng, Phys. Rev. B {\bf 98}, 245123 (2018).
\end{thebibliography}
\end{document}